\documentclass[aps,prl,reprint,groupedaddress]{revtex4-1}
\usepackage{graphicx}
\usepackage{dcolumn}
\usepackage{bm}
\usepackage[left]{lineno}

\begin{document}

\title{Physical model for turbulent friction on rough surfaces}

\author{Zhuoqun \surname{Li}}
\affiliation{Key Laboratory of Mechanics on Disaster and Environment in Western China, The Ministry of Education of China, Department of Mechanics, Lanzhou University, Lanzhou 730000, PR China}
\author{Xiaojing \surname{Zheng}}
\email[]{xjzheng@lzu.edu.cn}
\affiliation{Research Center for Applied Mechanics, School of Mechano-Electronic Engineering, Xidian University, Xi'an 710071, PR China}

\date{\today}

\begin{abstract}
We present a physical model for turbulent friction on rough surfaces with regularly distributed roughness elements. Wall shear stresses are expressed as functions of physical quantities. Surfaces with varying roughness densities and roughness elements with different aspect ratios are considered. We propose a straight forward method based on the conservation of momentum to deduce the drag on elements by expressing it as functions of the maximum drag and drag reductions ratios, as the drag on individual elements decreases as packing density increases. A drag reduction effect of momentum redistribution is proposed and the mutual sheltering effect is studied. These two drag reduction mechanisms for individual elements are significant for sparse and dense surfaces, respectively. Reduction ratios for redistribution effect and mutual sheltering effect are deduced, for the two different types of rough surfaces. The shear stress on elements and the total wall shear stress are obtained as the result of the drag analysis. The estimated wall shear stresses of the proposed model are consistent with classical experimental measurements.
\end{abstract}
\maketitle

\begin{figure}[b]
\includegraphics[width=0.5\textwidth]{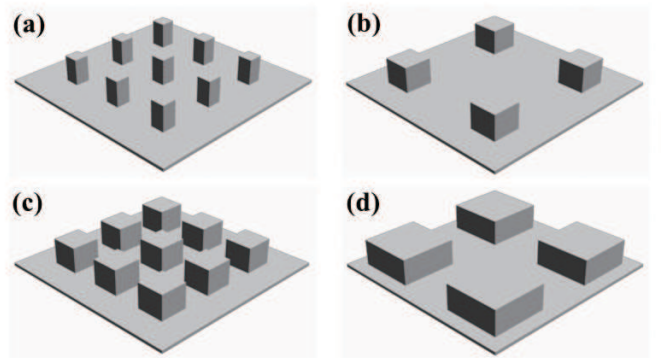}
\caption{\label{fig1}Four examples of rough surfaces with different configurations of $b/h = 1/2, 1, 1~\text{and}~2$; $b^2/D^2 = 1/16, 1/16, 1/4~\text{and}~1/4$; and $bh/D^2= 1/8, 1/16, 1/4~\text{and}~1/8$.}
\end {figure}

Turbulent flow over a rough surface is different from that over a smooth surface, as roughness elements disturb the flow and generally enhance turbulent friction. Quantification of turbulent friction over rough surfaces is the key to determine this disturbance and to characterize the velocity profile\cite{[{Translation in English: }]SchlichtingE, Jimenez2004, Smits2011}. It has been an open problem for boundary layer studies for decades. Due to the complexity of flow condition over rough surfaces, studies on this subject relied on empirical functions to approach measurements of wall shear stress. The pioneering experimental study of Nikuradse\cite{[{Translation in English: }] NikuradseE} parameterized surface roughness with a sand roughness height and found that the friction factor deviates from the turbulent smooth-wall law as Reynolds number $Re$ increases, while it becomes independent of $Re$ at higher $Re$\cite{NikuradseE}. However, another classical experiment of Schlichting\cite{SchlichtingE}, found that the drag on rough surfaces depends on both relative height of elements and roughness density even for large $Re$. This result implies that the height of element $h$, the breadth of element $b$ and the distance between the centers of adjacent elements $D$ are needed to parametrize a rough surface to account for possible dependencies of wall shear stress on the geometries of surfaces. Exemplary rough surfaces with different configurations are shown in Fig~\ref{fig1}. In this paper, we use packing density $b^2/D^2$ and aspect ratio $b/h$ to quantify surface geometries.
Schlichting pointed out that the total shear stress can be distinguished into the shear stress on rough surface and the shear stress on smooth surface, thus they can be measured separately and determined independently\cite{SchlichtingE}. A common definition of total wall shear stress $\tau$ is
\begin{equation}
\tau=\tau_r+\tau_s \label{t}
\end{equation}
here $\tau_r$ is the shear stress on roughness elements, $\tau_s$ is the shear stress on underlying surface\cite{Raupach1992}. Schlichting found that $\tau$ increases as roughness density increases, and approaches its maximum at certain roughness density. Perry\cite{Perry1969} carried out measurements of flow field over regular rough surfaces, and categorized rough surfaces into k-type and D-type, and described two distinctive mechanisms of turbulent friction working for these two types of surfaces, respectively\cite{Perry1969}. Marshall\cite{Marshall1971} then conducted an unique experiment which measured directly the drag on elements over surfaces with wide ranges of roughness densities and aspect ratios. In Marshall's experiment, the range of $b^2/D^2$ spans from 0.002 to 0.44, the aspect ratios are 0.5 to 5 and the height of elements is uniformly 2.54cm. The measured drag on individual elements decreases as $b^2/D^2$ increases. Based on Marshall's experiment, Raupach\cite{Raupach1992} deduced a drag partition theory for sparse surfaces with $b/h=1$, covering k-type surfaces, based on a hypothesis for the size of wake. In this drag partition theory, the ratio of $\tau_r/\tau_s$ is proportional to roughness density and it is validated against the experiment of Marshall\cite{Marshall1971}. Raupach's discovery encouraged us to believe that, although the details of non-linear behavior of turbulent flow over rough surfaces are difficult to observe and quantify, it is possible to parameterize the distribution of drag on surfaces with rational functions in association with the geometry of surfaces. However, hypotheses on the size of wakes are not validated by experimental measurements, existing theories do not directly estimate wall shear stress or drag on rough surfaces\cite{Raupach1992, Jimenez2004} and there are still problems concerning mutual sheltering effect for dense surface, i.e., D-type surface\cite{Yang2016}.

This paper studies turbulent friction on rough surfaces in a deep fully-developed turbulent boundary layer without obvious Reynolds number effect \cite{NikuradseE, Marshall1971, Raupach1992}, including both sparse surfaces and dense surfaces. On each surface, the elements as well as the flow around each element are considered identical showing periodic patterns. For this reason, each element and its occupied area $D^2$ of underlying surfaces are referred to as a unit surface. The flow in the roughness layer obeys the conservation of momentum over each unit surface. The flow above the roughness layer is considered homogeneous and the vertical momentum flux is a constant for all heights, related to the dimensions of the flow field, e.g., pipe diameter or boundary layer thickness, regardless of the geometry of surfaces. This constant vertical momentum flux determines the limit of total momentum that can be absorbed by the surfaces, which is also assumed as a constant for certain flow condition and defined at the top of roughness layer, in the form of shear stress as $\tau_{rr}$. Both the momentum of the flow in the roughness layer and the momentum absorbed by the surface contribute to $\tau_{rr}$, which could be considered as the total shear stress in fluid in the immediate vicinity of the wall \cite{Gioia2002}. When wall shear stress on a unit surface increases the momentum in the flow and the momentum flux that impact an individual element decreases, consequently the drag on each element is reduced, relatively comparing to the drag on isolated elements. This drag reduction effect for individual elements is named as redistribution effect. It occurs when $b^2/D^2$ increases. For dense surfaces, a mutual sheltering effect, which concerns the interaction among the wakes of elements and the main flow, is significant \cite{Jimenez2004}. Here we first discuss the drag and wall shear stress for sparse surfaces with the redistribution effect, before considering mutual sheltering effect for dense surfaces. 

\begin{figure}[b]
\includegraphics[width=0.5\textwidth]{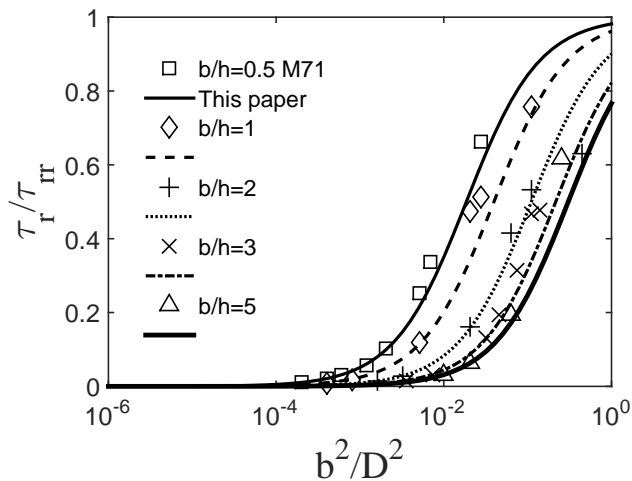}
\caption{\label{fig2} Comparison of wall shear stress on elements between the proposed theory and Marshall's data (M71).}
\end {figure}

\begin{figure}[t]
\includegraphics[width=0.5\textwidth]{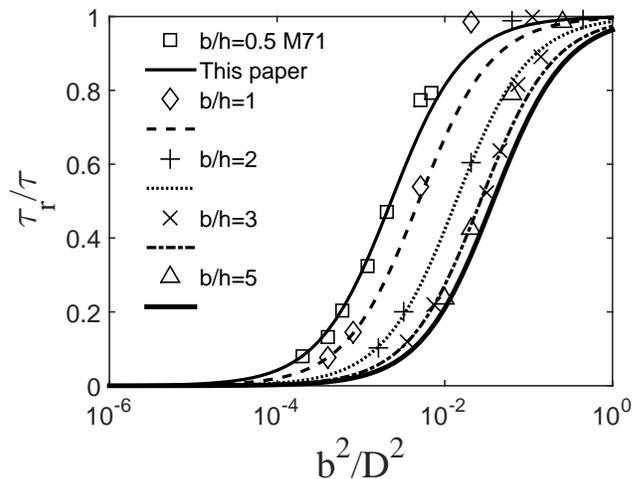}
\caption{\label{fig3} Comparison of drag partition between the proposed theory and Marshall's data (M71).}
\end {figure}

Drag on an isolated element is usually quantified by a drag coefficient $C_{r0}$:
\begin{equation}
C_{r0} = \frac{2w_{r0}}{\rho u^{2}_{h}bh}, \label{cf0}
\end{equation}
here $w_{r0}$ is the drag that acts on an isolated element, typically deduced from the measured shift of element under fluid pressure \cite{Marshall1971}; $w_{r0}$ is the maximum drag on an element under certain reference velocity for different $b^2/D^2$; $\rho$ is the density of air; $u_h$ is the reference fluid velocity at the height of elements, measured on smooth surface. The shear stress for an isolated element $\tau_{r0}$ is 
\begin{equation}
\tau_{r0}=\frac{w_{r0}}{D^2}. \label{tr0}
\end{equation}
Note that, for isolated elements, $\tau_{r} = \tau_{r0}$. Here also defines the shear stress on a smooth surface as $\tau_{s0}$, it is the maximum value of $\tau_s$ under certain reference velocity for different $b^2/D^2$.
The drag on elements $w_r$ and the drag on a unit area of underlying surface $\tau_s$ decrease, as $b^2/D^2$ increases, which is the result of the redistribution effect. As the redistribution effect influences both the drag on elements and underlying surfaces, here we propose a drag reduction rate $f$ as a factor in the expressions of $\tau_r$ and $\tau_s$. Let the momentum flux which impacts a unit surface being proportional to the shear stress on an unit surface, we have 

\begin{equation}
\left\{
\begin{array}{l}
\displaystyle
\tau_r=(1-f)\tau_{r0}, \label{tr}\\
\displaystyle
\tau_s=(1-f)\tau_{s0}, \label{tr}\\
\end{array}
\right.
\end{equation}
here $f$ is the drag reduction rate due to redistribution effect, it represents the ratio of the reduced amount of drag on an individual element to the drag on the isolated element. It also represents the ratio of the momentum flux that redistributed among all elements, i.e., $\tau_r$, to the total momentum flux $\tau_{rr}$, when the momentum flux is considered proportional to drag on a unit surface. Thus,
\begin{equation}
f=\frac{\tau_r}{\tau_{rr}}. \label{f}
\end{equation}
Note that, Eq.~\ref{tr} and Eq.~\ref{f} are valid for sparse surface where the mutual sheltering effect is weak and insignificant, and $\tau_{rr}$ is taken as the maximum of the momentum flux that can be absorbed by the elements. On denser surfaces, the estimated $\tau_{r}$ may not reach $\tau_{rr}$ as the mutual sheltering effect diverts and consumes momentum flux\cite{Jimenez2004, Yang2016}. Substituting $f$ in Eq.~\ref{tr} with Eq.~\ref{f}, we obtain the expressions for $\tau_{r}$ and $\tau_{s}$: 
\begin{equation}
\label{trts}
\left\{
\begin{array}{l}
\displaystyle
\tau_{r}=\frac{\tau_{rr}\tau_{r0}}{\tau_{rr}+\tau_{r0}},\\
\displaystyle
\tau_{s}=\frac{\tau_{s0}\tau_{rr}}{\tau_{rr}+\tau_{r0}}.\\
\end{array}
\right.
\end{equation}
Eq.~\ref{trts} are concise expressions for wall shear stresses on sparse surface. It can be used to calculate wall shear stresses with physical quantities. It can be also used to deduce momentum flux $\tau_{rr}$ with measurements of wall shear stresses. In another form of Eq.~\ref{trts}, wall shear stresses can be expressed as functions of parameters of surface geometry. Here we define a $w_{rr}$ as the drag on elements corresponding to $\tau_{rr}$. As the responds of elements with different $b^2$ to the momentum flux $\tau_{rr}$ are different, $w_{rr}$ is defined to be equal to $\tau_{rr}b^2$. Replacing shear stresses with drags and parameters of surface geometry in Eq.~\ref{trts} gives
\begin{equation}
\label{trtsw}
\left\{
\begin{array}{l}
\displaystyle
\tau_{r}=\frac{\tau_{rr}w_{r0}b^2/D^2}{w_{rr}+w_{r0}b^2/D^2},\\
\displaystyle
\tau_{s}=\frac{\tau_{s0}w_{rr}}{w_{rr}+w_{r0}b^2/D^2}.\\
\end{array}
\right.
\end{equation}
Eq.~\ref{trtsw} shows that $\tau_{r}$ and $\tau_{s}$ are functions of $b^2/D^2$. With a given aspect ratio, the dependencies of $\tau_{r}$ and $\tau_{s}$ on $bh/D^2$ can also be determined. Then the partition of drag is 
\begin{equation}
\label{tpw}
\frac{\tau_{r}}{\tau}=\frac{\tau_{rr}w_{r0}b^2/D^2}{\tau_{rr}w_{r0}b^2/D^2+\tau_{s0}w_{rr}}.\\
\end{equation}
With Eq.~\ref{cf0} and data from the experiment of Marshall \cite{Marshall1971}(see Tab.~\ref{tab1}, $\tau_{rr} = 6 Pa$ by Eq.~\ref{trtsw}), $\tau_r/\tau_{rr}$ as well as drag partition $\tau_r/\tau$ are calculated by Eq.~\ref{trtsw} and Eq.~\ref{tpw}, they are shown in Fig.~\ref{fig2} and Fig.~\ref{fig3}, respectively. The estimations and experimental measurements agree well for varying $b^2/D^2$ and $b/h$ in these two Figures.


\begin{table}
\caption{\label{tab1}Original data from wind tunnel measurements \cite{Marshall1971}, with $\tau_{s0} = 0.74 Pa$ and $u_h=20.3m/s$. }
\begin{ruledtabular}
\begin{tabular}{cccccc}
$b/h$ & 0.5 & 1 & 2 & 3 & 5\\
$C_{f0}$ & 0.64 & 0.6 & 0.43 & 0.33 & 0.39\\
$w_{r0} (N)$ & 0.0521 & 0.0977 & 0.140 & 0.1612 & 0.3175\\
\end{tabular}
\end{ruledtabular}
\end{table}
\begin{figure}[b]
\includegraphics[width=0.5\textwidth]{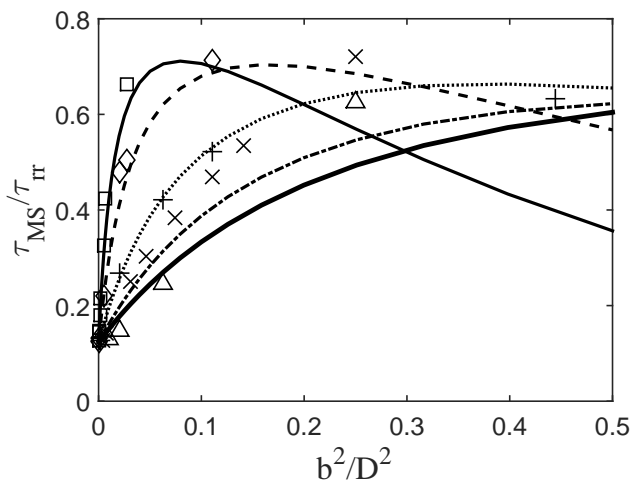}
\caption{\label{fig4} Comparison of $\tau_{MS}$ between the proposed theory and Marshall's data. The legend is the same as Fig.~\ref{fig2} and Fig.~\ref{fig3}. }
\end {figure}


Then we consider dense surfaces. It is generally believed that $\tau_{r}$ increases monotonically with $b^2/D^2$ before it approaches its maximum, as it is shown in Fig.~\ref{fig2}. And some measurements show a peak of shear stress at a threshold $b^2/D^2$, before $b^2/D^2$ reaches its maximum, e.g., Schlichting \cite{SchlichtingE} observed a peak at $bh/D^2=0.316$ (denser than Fig.~\ref{fig1}(c)) on a surface with spherical elements with a diameter of 4.1mm, out of 21 different types of surfaces. For dense surfaces, when it is almost fully packed with elements, $\tau$ is inevitably reduced to a value close to $\tau_{s0}$, if there is no gaps on the fully packed surface and $h$ is not significant comparing to the dimension of flow field \cite{Jimenez2004}. The mutual sheltering effect is assumed to be responsible for this reduction of shear stress. 

When an element is placed on a rough surface, the element expels a volume of fluid to the main flow. A portion of the momentum carried by the displaced fluid is absorbed by the surface, the rest is ejected upward to the main flow, and a wake is formed as the result of the development of the displaced fluid. When many more elements are placed on a surface, their wakes start to shelter each other and wall shear stresses are reduced, this is the mutual sheltering effect. The latest study on this effect is an empirical model based on simulations by Ref.~\onlinecite{Yang2016}. However, there is still empirical parameters due to difficulties in determination of the size of wake, thus we choose to study mutual sheltering effect by analyzing the momentum of displaced flow rather than the size of its developed form. 

As the direction of the displaced fluid and the momentum flux from the inertial layer to the surface are opposite, the total momentum flux and the shear stress on elements decrease as each element is added on the surface. In consistent with the definition of drag coefficient in Eq.~\ref{cf0}, we set the integrated momentum of the fluid displaced by an element as $0.5\rho u^{2}_{h}bh$. And the potential integrated momentum that could be displaced when the surface is fully packed is $0.5\rho u^{2}_{h}nD^2$, here $n$ is the number of elements in the whole flow field. Thus, the reduction rate of momentum flux and drag is $bh/(nD^2)$. The modified total wall shear stress by considering mutual sheltering effect is
\begin{equation}
\tau_{MS}=\tau (1-\frac{bh}{nD^2})^{n}, \label{tms}
\end{equation}
here $\tau_{MS}$ is the total shear stress concerning mutual sheltering effect. Eq.~\ref{tms} shows that the mutual sheltering effect is weak for small $b^2/D^2$, as estimated. When $n$ approaches infinity great, which means very dense surface, $(1-bh/(nD^2))^n$ approaches $e^{(-bh/D^2)}$. Thus, with $\tau$ from Eq.~\ref{trtsw}, $\tau_{MS}$ is 
\begin{equation}
\tau_{MS}=\frac{{\tau_{rr}w_{r0}b^2/D^2+\tau_{s0}w_{rr}}}{w_{rr}+w_{r0}b^2/D^2}e^{(-bh/D^2)}. \label{tsteee}
\end{equation}
Estimation of $\tau_{MS}/\tau_{rr}$ as well as the nondimensionalized total wall shear stress from Marshall's experiment\cite{Marshall1971} are shown in Fig.~\ref{fig4}. Most estimations in Fig.~\ref{fig4} are consistent with the measurements, the rest have discrepancies mainly from surfaces with large aspect ratios. The cause of discrepancies is the cylindrical elements used in Marshall's experiment, which means there are still gaps when the surface is fully packed. We tried to replace $b^2$ with $0.25\pi b^2$ to account for the cylinders in the raw data, but such modification does not eliminate the discrepancies. The influence of differences in shapes of elements to the drag law remains to be discovered. However, Fig.~\ref{fig4} as well as Fig.~\ref{fig2} and Fig.~\ref{fig3} are satisfactory results for a general physical model of wall shear stress. Note that, Fig.~\ref{fig4} cannot serve as a validation for the existence of the peak of total shear stress, we need more drag data to validate Eq.~\ref{tsteee} on very dense surfaces with elements of large aspect ratios, particularly direct measurements of drag on elements. 

In the proposed model for turbulent friction, the expressions for wall shear stresses, with or without mutual sheltering effect, are analytical expressions of physical quantities. The mechanisms of drag reduction on individual elements due to redistribution effect and mutual sheltering effect are considered, for sparse and dense surfaces. Both reduction rates of these two mechanisms are expressed as factors in the model. This model is applicable for surfaces with varying $b^2/D^2$ and $b/h$. Drag on surfaces determines the structure of flow near surfaces, particularly for sparse surface. In the model, $\tau_{rr}$ serves as the link between the shear stress on surfaces and the momentum flux in the inertial layer. It can be deduced by measurements of wall shear stresses, yet difficult to measure directly or simulate with DNS at high $Re$. Studies are required to associate $\tau_{rr}$ with parameters of velocity profile by the proposed model, to determine the contribution of roughness to the structure of flow. 

The key to success in building this model for turbulent friction is the decomposition of this problem. The rough surfaces are firstly distinguished into sparse and dense surfaces and discussed respectively. Then the two drag reduction effects for individual elements are proposed and quantified independently. And most importantly, drags and shear stresses are treated differently as their dependencies on parameters of the geometry of surfaces are different. This method allow us to simplify a non-linear process into independent linear relationships among physical quantities and form a physical model. Also for this reason, the estimated quantities are analyzed and determined independently and only nondimensionalized when they are plotted in the diagrams. There could be different ways to understand the proposed expressions and mechanisms of momentum transfer in the roughness layer. However, the most urgent practice is to acquire more data to validate this model for very dense surfaces and locate the peak of wall shear stress.

\begin{acknowledgments}
This work is financially supported by NSFC projects (Nos. 11490553, 11232006 and 11121202).
\end{acknowledgments}
\providecommand{\noopsort}[1]{}\providecommand{\singleletter}[1]{#1}%
\end{document}